\documentclass[12pt,preprint]{aastex}

\slugcomment{accepted by PASP, October 2006 issue} 

\shorttitle{stare-mode astrometric mission}
\shortauthors{Zacharias \& Dorland}

\begin{document}

\title{The concept of a stare-mode astrometric space mission}


\author{N.~Zacharias and B.~Dorland}
\affil{U.S.~Naval Observatory, Washington, DC 20392,
      \email{nz@usno.navy.mil, bdorland@usno.navy.mil} }

\begin{abstract}
The concept of a stare-mode astrometric space mission is introduced.
The traditionally accepted mode of operation for a mapping astrometric
space mission is that of a continuously scanning satellite, like
the successful Hipparcos and planned Gaia missions.
With the advent of astrometry missions mapping out stars to 20th
magnitude, the stare-mode becomes competitive.
A stare-mode of operation has several advantages over a scanning
missions if absolute parallax and throughput issues can be 
successfully addressed.
Requirements for a stare-mode operation are outlined.
The mission precision for a stare-mode astrometric mission is derived 
as a function of instrumental parameters with examples given.
The stare-mode concept has been accepted as baseline 
for the NASA roadmap study of the Origins Billions Star Survey
(OBSS) mission and the Milli-arcsecond Pathfinder Survey
(MAPS) micro-satellite proposed project.
\end{abstract}



\section{Introduction}

The current paradigm for a mapping, astrometric space mission is a 
scanning satellite with 2 fields of view (FOV) which are separated 
by a large angle of $\approx$ 50$^{\circ}$ to 100$^{\circ}$.
This concept was first introduced by P.~Lacroute \cite{lacroute}.
The 2 FOV are imaged onto the same focal plane, thus relative 
position measures of stellar images in the focal plane allow 
wide-angle measures on the sky.
This observing strategy leads to a well conditioned least-squares 
problem to solve for absolute parallaxes.
For any small area in the sky the parallax factor \citep{vandekamp}
is about the same, leading to large correlations in a
global solution for absolute parallaxes based on stellar, all-sky 
observations of small angle astrometric measures alone.

This concept of 2 FOV separated by a ``basic angle" worked very well 
for the ESA Hipparcos mission \citep{hip}.  This concept has been 
adopted also for the planned ESA Gaia mission \citep{gaia}, as well 
as the canceled FAME mission \citep{fame} and the unfunded 
DIVA \citep{diva} and AMEX \citep{amex} missions.
Another advantage of this concept is the almost 100\% efficiency
in data collection with continuous observations in time-delayed
integration (TDI) mode using charge-coupled devices (CCDs).
Together with the success of the Hipparcos mission, 
the scanning concept has been adopted as ``the optimal" 
concept for a mapping astrometric mission.\footnote{The Space 
Interferometry Mission (SIM) is also a dedicated,
astrometric mission; however, it operates on a totally different
concept (interferometry) and can observe only a very limited
number of pre-selected targets.}
However, going to much higher accuracies and fainter limiting
magnitudes with access to galaxies and quasars justifies a
second look at the basic operation principle and possible
alternatives.  

The stare-mode concept for an all-sky, mapping, astrometric
space mission is presented here, which requires only a single
FOV and operates differentially.  This is contrary to the quasi-absolute,
large-angle measure principle using 2 FOV with a scanning mission.
A stare-mode operation with 2 FOV is technically feasible.
However, this approach would avoid only some of the issues
raised here for a scanning mission and is not discussed further.

In section 2 the disadvantages of scanning, astrometric
space missions are presented and in section 3 the stare-mode
concept is presented as an alternative to overcome these problems, 
stressing the requirements which need to be meet in order to be a 
viable option.
In section 4 mission precision and other relevant mission 
parameters are derived from instrumental and basic input
parameters.  
Section 5 discusses some realistic examples for small and large-scale 
astrometric missions based on this stare-mode concept.
The Origins Billions Star Survey, OBSS \citep{obss} study adopted as
baseline the proposed stare-mode operation principle.
OBSS is a NASA sponsored investigation for NASA roadmap planning.
The OBSS study report was submitted in May 2005 \citep{obssr},
and more details are given elsewhere \cite{obss_science}.

\section{Disadvantages of a scanning astrometric mission}

The main issues with a scanning astrometric mission
like Hipparcos and Gaia are:

\begin{description}
\item[Basic angle stability.]
In the 1 mas regime (Hipparcos) the basic angle stability was
already technologically challenging.  For 10 $\mu$as this
becomes a major cost driver and primary source of systematic
errors.  For example the Gaia design \citep{gaia} requires 
passive temperature stability in the micro-Kelvin regime and 
an active metrology system.

\item[2 apertures separated by large angle.]
This has the advantage of direct, wide-angle measurements,
but on the other hand becomes a problem with respect to image 
confusion and crowding in the focal plane and instrument complexity 
with the tough engineering requirement of having no significant
beam walk to make use of this concept as intended.
The complexity of a beam-splitter hardware is costly.

\item[CCD vs.~CMOS.]
The scanning concept relies on driving focal plane detectors
in TDI mode, which can be accomplished only with CCDs.  
Complementary Metal Oxide Semiconductor (CMOS) detectors,
(or hybrid detectors) which show promise for future applications, 
may be better suited for space applications, being radiation hard.
In addition, CMOS pixels don't spill
charge (blooming) and thus are inherently better suited
for bright star astrometric observations and for spanning
a very large dynamic range, even when small pixel sizes are 
considered.
Current CCD technology supports various anti-blooming features.
However, the most effective, lateral anti-blooming becomes 
increasingly problematic for CCDs as the pixel size decreases, 
effectively becoming impractical at the 8 $\mu$m size.

\item[Scanning restrictions.]
The scanning mission is limited to a specific scanning law.
No target of opportunity can be observed.  Also, the integration
time is fixed (optimized for uniform scanning speed), and 
typically relatively short (i.e.~a few seconds) 
due to other constraints, forcing the mission
to use a large aperture to reach faint limiting magnitudes.  
Slower scanning is undesirable for satellite stability 
and other reasons.  Large apertures drive focal length,
mass and cost and make access to bright stars problematic.
The scanning law and the Sun exclusion angle lead to an uneven
sky coverage with a typical average mission precision variation
of astrometric parameters by a factor of 2 as a function
of ecliptic latitude.  
No specific target areas can be observed with other precision
than dictated by the scanning law.  The temporal cadence of
observations has no flexibility.

\item[Image smearing.]
The spacecraft angular momentum vector needs to precess to 
produce an all-sky survey.  The continuous precession results 
in image smearing.  Remaining differential distortions
over the field of view add to image smearing.
Elongated, or generally asymmetric image profiles increase the 
astrometric errors, random as well as systematic errors.

\item[Jitter.]
Small non-uniformities of the scanning (spacecraft jitter) cause
changes in the image profiles as a function of time.  The TDI
mode does not integrate all stars in a given field simultaneously,
thus different stars observed almost at the same epoch (same FOV), 
are affected differently, leading to positional offsets which
need to be modeled or cause additional random and systematic
errors.

\item[1-dim data.]
The scanning operation gives only 1-dimensional measures of
high precision.
This has some advantages (i.e.~simple profile fit) but has significant
disadvantages in the later stages of the data reduction and global
astrometric re-construction (error propagation issues, mixing with 
instrumental effects, attitude control).
Also for many applications (parallaxes, planet detections) the
1-dim observations will be quite often (near 50\%) along the
``wrong axis", while 2-dim data gives results for any
projection angle for any single observation.

\item[Downlink.]
It is difficult to use a steerable, high gain antenna (HGA) to 
achieve a large downlink rate from an L2 orbit.  A spinning 
satellite, in order to achieve comparable data rates ($\approx$ 
30 Mbps), must either be positioned close to Earth or employ a 
dedicated relay satellite equipped with a direction HGA at 
significant additional cost to the program.
A stare-mode satellite could be equipped with a HGA.

\end{description}

\section{The stare-mode concept}

For a stare-mode mission (with a single FOV) to be considered as 
a viable alternative to the scanning satellite concept, 
2 basic issues need to be addressed and solved:

\begin{enumerate}
\item Global astrometric accuracy from stitching together
  small, overlapping fields, without the advantage of direct
  large angle measurements (particularly for absolute
  parallax determination).
\item Overhead time (observing efficiency) for setting to the 
  next field needs to be short including actual slew of spacecraft, 
  settling time, guide star acquisition and detector read out time.
\end{enumerate}

\subsection{Global astrometry with block adjustment techniques}

\subsubsection{Traditional, ground-based technique}

The stare-mode idea presented here follows closely the traditional, 
photographic astrometry survey principle.  The telescope is pointed 
at a field of view and all stars in the focal plane are integrated 
simultaneously while the pointing and field orientation angle are 
kept constant with the help of 2 or more guide stars.  
Then the FOV is shifted and the next field is exposed.  
Consecutive FOV are overlapped by typically
25 to 50\% in area.  A zonal pattern of the sky is covered within
a short period of time which is eventually supplemented by similar
adjacent, overlapping observations to cover the entire sky.

This pattern of overlapping fields allow for a block adjustment 
(BA) 
(Eichhorn 1960; de Vegt \& Ebner 1974; Googe, Eichhorn \& Lukac 1970),
in which the astrometric and instrumental parameters (``plate constants") 
are estimated at the same time in a single, rigorous, 
non-linear, least-squares adjustment.
Any applicable reference star catalog will be sufficient for an
initial reduction of the data to get approximate starting values 
for the linearized BA procedure, which converges typically in a single
iteration step.  
The problem is rank deficient; an external orientation of the
global coordinate system needs to be provided\footnote{The same is
true also for a scanning mission.}.
Usually the then available ``best" celestial reference frame is 
chosen for this external orientation to be consistent with the 
previous realization of such a system.

Alternatively to the BA reductions an iterative conventional adjustment 
(ICA) scheme can be used to perform the global reductions.
A classical ``single plate" adjustment gives improved positions for 
reference and selected field stars on a frame-by-frame basis.
For individual stars then data are combined to obtain mean position, 
proper motion and parallax, from all overlapping fields and from 
different epochs.  These improved data are fed into the next 
iteration to repeat the adjustment.
The ICA scheme converges to a consistent, global catalog up to the 
accuracy limits of the input $x,y$-coordinates on a reference system 
which is represented by the average of all the original reference 
catalog star coordinates (external system orientation and rotation).
Both the BA and ICA approach give equivalent results \cite{BST}.
The BA approach is conceptually ``cleaner" than the ICA but
takes a hugh amount of computer resources.
Contrary to just the single-step classical ``plate reduction",
the BA and ICA concepts explicitely utilize the astrometric information
(same star on different exposures has only 1 set of astrometric
parameters) from overlapping fields, involving all (suitable)
stars in the reductions, not just the few reference stars.

The BA concept has been successfully applied for a few cases
\citep{agk2_ba, cpc2_ba} and has been studied with simulations
\citep{ba_nz}. 
Important to realize is that with a BA type reduction the
scale and orientation (roll angle) parameters of each
individual exposure is determined very precisely.
If there is for some reason a jump in scale between exposures
or a very small drift, it will not affect the astrometric
results of the reductions.  Scale and orientation are 
overall very few parameters in a well conditioned system
of observation equations when dealing with many star images
per exposure.

Although these applications so far dealt with positions only,
the formal extension of the principle and algorithm to include 
proper motions and parallaxes is straightforward.
However, deriving {\em absolute} parallaxes and proper motions from 
small angle measures (ICA or BA) deserves some elaboration (see below).

\subsubsection{Difference between SIM and astrograph-type observations}

The Space Interferometry Mission (SIM) \citep{sim} has only
a single field of regard (FOR); however, this FOR is relatively large
(15$^{\circ}$).  Rigorous simulations \\  \citep{simsim} have shown
that the mission goals for all 5 astrometric parameters (position,
proper motion and parallax) can be achieved, although the 
least-squares system is not well conditioned, at least if just 
stars are used for the astrometric grid as originally planned.
Recent simulations \cite{simDDA} show that by including even a
small number of extragalactic, fixed fiducial points ($\approx$ 25
QSOs) the absolute errors in parallax become significantly smaller
(about factor of 2).

It is important to keep in mind that SIM observations are fundamentally
different than the astrograph-type mapping observations suggested here.
SIM does obtain (relatively) large-angle, absolute measures. 
However, SIM observes only 1-dimensional, angular separations between 
2 targets at a time and a global grid needs to be stiched together
by these quasi absolute angle measures, including all instrumental
parameters before any catalog of positions and motions can be established.
A single astrograph-type observation yields differential, 2-dimensional
positions for thousands of targets simultaneously with a minimal
number of instrumental parameters.
Relative proper motions can be derived from astrograph-type 
observations of the same field at 2 or more different epochs, 
even without any overlap to adjacent fields, see e.g.~the 
Northern and Southern Proper Motion Surveys, NPM, SPM \cite{npm, spm}.

\subsubsection{Will the stare-mode reductions work for global, 
               astrometric, space missions?}

Block adjustment procedures in ground-based, traditional, 
photographic astrometry can successfully be applied 
even in the case of very few reference objects \citep{ba_nz}.  
The BA technique does not work very well for a zonal pattern (fields 
along a narrow strip around the sky) in the presence of systematic
errors, but is well conditioned for a hemisphere or all-sky 
coverage due to the many inherent ``closure conditions" 
(after going around the sky in a circle,
the same stars are mapped, forcing to the same positions,
give or take parallax/PM effects).
A narrow zone has closure only along 1 axis, while the
hemisphere or all-sky case is much ''stiffer" with closures
in 2 dimensions.  Imagine a zone of few degrees width around
the equator.  Coordinates along RA are well constraint due
to the closure at 0/24 hours RA.  However the star positions
could easily have systematic errors along declination, for
example as a function of RA.  With the entire hemisphere
covered, there are multiple constraints reaching from one
side over the pole to the other side to "fix" zonal errors.

Critical for the success of a BA or ICA approach, besides
a hemisphere or all-sky coverage are 4 issues:

\begin{enumerate}
\item Sufficient overlap between adjacent fields.
\item Sufficient number of link stars in overlapping frames.
\item Fiducial points for absolute parallax and proper motion 
  determination.
\item High instrumental stability of 
  {\em higher order} variations (mapping model).
\end{enumerate}

The first item is easily accommodated. The stare-mode operation
allows for a flexible cadence to observe adjacent fields with 
overlaps as required.

The second item requires a large number of stars per unit area
in the sky.  This typically becomes feasible with a faint limiting
magnitude of the instrument.  For a Hipparcos-type mission this
actually would have been a problem but it is not an issue for
a Gaia or OBSS type mission.  
For a well conditioned system, the mission precision, $\sigma_{m}$
(mean astrometric errors for a given target object, averaged over all
observations of that target during the lifetime of the mission) will
approach the $\sigma_{m} = \sigma_{smp} / \sqrt{n_{t}}$ limit, with the
single measure error $\sigma_{smp}$ and the number of observations,
$n_{t}$, per target (per coordinate).

Simulations with only a 4-fold overlap pattern of a hemisphere
using a small FOV of $4^{\circ}$ and only about 200,000 stars
lead to a well conditioned system with the actual $\sigma_{m}$ 
being larger than the square-root-n limit by only about a factor 
of 1.04 \citep{ba_nz} for star positions.

Random errors of individual stellar position observations can lead to
a zero-point offset of an entire field of about 
$\sigma_{smp} / \sqrt{n_{s}}$, with $n_{s}$ being the number of (well exposed)
stars in that field of view.  If $n_{s}$ is significantly larger
than $n_{t}$, this zero-point offset of a field (i.e.~zonal error)
is likely to be smaller than the envisioned mean mission errors.
In the example cases given below this requirement is met,
and the BA reduction procedure likely will give the expected 
performance.  Detailed simulations with the specific mission
parameters for the OBSS case are planned for a phase A study.

The third requirement can be met by observations of quasars and
compact galaxies.  As soon as the limiting magnitude of the instrument
can access a significant number of these extragalactic sources they
provide absolute reference points for parallaxes and proper motions.
This concept is proven even for not-much-overlapping, differential, 
ground-based observations, for example by the Northern and Southern 
Proper Motion projects, NPM and SPM \citep{npm, spm}.
Again, this was not an option for Hipparcos but is not an issue
for either OBSS or Gaia, which reach limiting magnitudes of 
21 and 20, respectively.
The zone of avoidance, i.e.~the galactic plane with high extinction
areas is a comparatively small area in the sky for a global program
and the BA technique of linking all FOV should be able to ``bridge"
those areas (pending further simulations to verify this assumption).
Furthermore, some optical counterparts of quasars have been
confirmed at very low galactic latitudes \citep{rorf}, and
only a small number of fiducial points are required to set the
zero-point for absolute parallaxes and proper motions.

The last item needs engineering attention in any case, 1 or 2
FOV, scanning or stare-mode observations.  However, for the
2 FOV, large-angle measure approach, challenging basic-angle 
stability/monitoring requirements have to be added.
The stare-mode option has the big advantage of being totally
{\em differential}.  Even a change in scale from one FOV to
the next is easy to handle; only a few instrumental parameters
like zero-point, scale and orientation per FOV contrast the
large number of observations (individual x,y data of stellar 
images observed simultaneously) and the large number of overlap 
connections (adjacent fields, number of repeats of all-sky pattern).
This leads to relatively low requirements on
thermal gradients etc.~in the instrument design with significant
cost benefits as compared to a quasi-absolute, large-angle
measure approach.  The only requirement is that {\em higher
order} mapping terms (field distortion pattern etc.) are
``calibrated out" and a simple mapping model must describe the
individual frames in the final BA (after an iteration and
calibration).
If too many parameters and changing calibration values over
short periods of time are required, the global astrometry would 
suffer from significant error propagation losses.

\subsection{Overhead}

The stare-mode operation is potentially more inefficient than 
a spinning observatory that employs TDI-mode observing.
There are 2 primary sources of inefficiency:
first, during step-stare observing, the detectors must be read out.
The readout period must be sufficiently long to permit low read
noise from the detector amplifiers and the readout electronics.
Using standard CCD technology, no photons can be collected during readout,
making the readout period essentially dead time in terms of integration.
If, on the other hand, active pixel sensor (APS) detector technology such as
CMOS or CMOS-Hybrid sensors are used, pixels are read out while integration
continues, eliminating this source of inefficiency.  APS detector technology
has the added benefit of supporting electronic shuttering, eliminating the
need for a mechanical shutter.

The second source of overhead is repositioning of the field of view
after a field has been observed.  During the repositioning of the
field of view (and during any subsequent settling period of the
structure), astrometric observations cannot be taken.
The overhead can be minimized by overlapping the read-out time
with the reposition and settle time.
However, it is important to remember that a scanning mission 
observes only 1-dim data while the stare-mode approach obtains
2-dim data simultaneously, thus starting out with a factor of
2 advantage during the time of photon collection.

A somewhat lower efficiency is not necessarily a bad thing for 
astrometry, if a higher astrometric quality (smaller systematic
errors) is obtained.
All dedicated astrometric instruments lose photons.
Hipparcos utilized a modulating grid in the focal
plane to observe a star at a time, disregarding all the other
targets which would have been accessible simultaneously.
Astrographs use narrow filters and sometimes grating images,
reducing the limiting magnitude.
Note, the spinning observatory loses efficiency in other places:
unfavorable error propagations with 1-dim observations
when scans do not intersect near orthogonal and 1-dim observations
at a time.

Assuming an operation principle similar to HST which takes minutes 
to re-position to the next FOV, the stare-mode astrometric mission 
would not be an option.  
Current technology provides a couple of different solutions that
lead to acceptable overhead times.  For small apertures ($\le$ 0.5 m)
one can envision a moveable, full-aperture scan mirror
that, for example, rotates in half-degree increments so as to
access a large arc (100 to 360 degrees) for a step-stare instrument 
without requiring reorientation of the observatory.
Such an instrument has been discussed and appears feasible
(K.~Aaron, priv. comm.).  As the aperture increases, however,
the size and mass of the support structure, and method of moving the 
full aperture flat scanning mirror become increasingly problematic.
At 1.5 m, a 45 degree inclination of the scan mirror with respect
to the optical axis would result in a 2.2 m flat.  This optical
element must be properly stowed in the available fairing, must be
deployed on-orbit, the torsional effects on the optical structure
due to rotation of the mirror must be compensated, and a non-trivial
mechanism deployed in order to move the large flat.
Moving the entire observatory seems more practical at least for large
aperture stare-mode missions, and likely even for small apertures
\cite{maps}. 
This would not be feasible using reaction wheel technology, 
with slew and settle times of order 100 seconds per field, 
and a resultant $\approx$ 10\% observing efficiency.
Control Moment Gyros (CMGs), or constant speed wheels, provide much
higher torques, have proved space heritage, and are commercial,
off-the-shelf items.

A dead-time (no photon collection) of about 50\% of the time over
the entire mission can be considered as very efficient as
compared to a scanning operation (see above 1-dim versus 2-dim observations).
The number of collected photons, leaving all other mission parameters
the same would be less in the stare-mode option, but the
number of individual observations (stellar image coordinates) would be 
the same for both cases.

It should be noted that the stare-mode concept, with its constant
readout and slew times, is naturally more efficient when
taking long exposures.  A 100 sec exposure with a 10 sec 
overhead results in 90\% observing efficiency, for example. 
When combined with the fact that longer exposures using similar 
optics and detector performance result in a fainter noise floor, 
the stare-mode observing mode would appear to be more naturally 
suited to fewer, deeper exposures vs.~the scanning observing mode.

The loss due to overhead time in a stare-mode mission can be
compensated by increasing the size of the focal plane array
(see also section 4), at the expense of a moderately reduced aperture.
The stare-mode still can go deeper (longer integration time)
and is not restricted by the downlink rate (directed beam antenna
is possible) as e.g.~the FAME mission concept was.
With selected field observing, possible if desired in stare-mode, 
denser areas can be targeted more often which dramatically enhances 
the ratio of observing stars versus observing empty space.

\subsection{Stare-mode operational details}

Operation of the envisioned stare-mode astrometric mission would
go as follows.  The telescope slews to a new field, locks on at
least 2 guide stars, settles and starts tracking.
Then the longer of 2 exposures begins, followed by a read-out
of the detector array, while the pointing of the telescope is
still fixed with guide trackers running.
The second, shorter exposure is taken.
While the detector array reads out again, the guiders are
disengaged and the telescope moves to the next, adjacent field.

The 2 exposure times of different duration (e.g.~factor of 10)
allow coverage of a large dynamic range and also are good
astrometric practice to check on possible magnitude-dependent
systematic errors.  More than 2 exposures per pointing can
be observed if required, e.g.~with different filters, if
a design including refractive elements is favored.

On-board processing would include the raw data processing
(bias and flat field corrections) as well as object detection.
The expected cosmic ray environment may necessitate
on-board, first order cosmic ray rejection logic.
Pixel data from small sub-areas around each detected object 
are saved, compressed and eventually relayed to the ground
together with instrumental parameters and knowledge of the
field position in the sky, time and integration time.

In order to apply the BA or ICA reduction techniques, adjacent fields
need to be observed with significant overlap in sky area.
In order to solve for parallaxes and proper motion, several
all-sky coverages per year need to be completed.
Examples (see below) show that neither requirement poses a 
problem.

The actual observing cadence can be very flexible, with
for example longer integration times and more data taken
in selected areas of scientific interest or faint targets. 
Spending on the order of a few hours of observing time on
a small area in the sky could result in quasi single epoch
mean positions on the microarcsecond precision level
(see 1.5 meter telescope example in section 5).
Another option is to observe specific fields very often
throughout the mission for good temporal sampling.
The general all-sky survey can be made uniform if
desired with the same number of observations per field
all over the sky.

A mix of an all-sky survey and targeted observing mode
is likely to give extraordinary scientific return
\citep{obssr, obss_science}.
Part of the total observing time is spent on an all-sky 
survey spread over the entire mission life time to be able to
solve for absolute parallax and proper motions with high accuracy.
The remaining observing time is allocated to specific
target areas with observing cadence and limiting magnitude
tailored to specific science goals.

\section{Mission precision}

\subsection{Definitions and assumptions}

How good can a stare-mode astrometric mission be?
In the following discussion the ``best mission precision", $\sigma_{m}$, 
is derived from some basic mission parameters and assumptions.
In general astrometric precision will depend on the brightness
of the stars.  For saturated, overexposed stars no high precision results
are assumed, while for faint stars the precision also drops due to the 
low signal-to-noise (S/N) ratio.  Thus the best astrometric precision
is achieved for stars near but just below the saturation limit.
The width of this ``sweet spot" depends on how much the overall error
is dominated by just the S/N or other errors (see below).

The following assumptions are made.

\begin{enumerate}
\item The location for the best astrometry ``sweet spot" on the magnitude 
  scale is not forced to a particular value.  There is no requirement for 
  a certain positional error at a certain magnitude. 
  The results obtained below will need to be shifted to
  a desired magnitude interval by additional considerations.  
  Particularly, enough overlap stars need to be available which
  excludes scaling to very small apertures and low limiting magnitudes
  with a small field of view (see examples and discussion below).
  The goal here is to find the smallest astrometric error overall,
  regardless of at what magnitude that might be.
\item Consider mainly random errors, thus deal with precision, not accuracy.
  However, to be realistic, a systematic error floor level is assumed 
  for the single measure precision independent of photon statistics.
\item Assume a well-conditioned system with nearly no ``loss" due to
  error propagation.  Thus the mean precision of a star 
  position after a completed mission follows the square-root-n law.
\end{enumerate}

\subsection{Calculations}

For the purpose at hand let the instrument and mission be defined 
by the set of basic input parameters given in Table 1.
In particular, the single measure precision, $\sigma_{smp}$,
is the assumed error floor, a combination of random and systematic
errors on the individual stellar image centroiding level,
and given as a fraction of a pixel.
The $\sigma_{smp}$ will depend on the astrometric quality of
the hardware and thus be different from case to case as a function of 
many technical details of the telescope, detector and operations.
In the examples to follow in section 5 realistic values are introduced 
for this free parameter. 
Astrometric quality is explained in more detail in a design study
of the USNO Robotic Astrometric Telescope  \citep{urat}.
Furthermore in the algorithm to follow a circular aperture and focal 
plane is assumed, however any shape leads to the same conclusions.

Table 2 summarizes the quantities derived from the set of basic
parameters and also shows the equations and units used.
The goal is to arrive at the overall mission precision, $\sigma_{m}$, 
for a single, stellar position coordinate of a star within the
``sweet spot" of brightness.

Using the algorithm of Table 2, quantities are then back substituted
into the mission precision, $\sigma_{m}$ equation to allow only
basic input parameters (Table 1).
Dropping all numerical scale factors (and some unit conversion factors)
gives the following proportionality equations:

\[ n \ = \ T \ \nu \ n_{ex} \ \sim \ T \ n_{ex} \ A \ / \ t_{p} \]

\[ n \ \sim \ T \ n_{ex} \ d^{2} \ s^{2} \ / \ t_{p} \]

\[ n \ \sim \ \frac{T n_{ex}}{t_{p}} (\frac{d}{f})^{2}  \]

\[ \sigma_{m} \ \sim \ \sigma_{smp} \ p_{s} \ / \ \sqrt{n} \]

\[ \sigma_{m} \ \sim \ \sigma_{smp} \ \frac{FWHM}{S} \ 
   \sqrt{\frac{t_{p}}{T n_{ex}}} \ \frac{f}{d} \]

\[ \sigma_{m} \ \sim \ \sigma_{smp} \ \sqrt{\frac{t_{p}}{T n_{ex}}} 
   \ \frac{f}{d S} \ \frac{\lambda}{a}  \]

This is the best astrometry (``sweet spot") achievable as a
function of the basic parameters as defined above.
A few, newly defined key items simplify this equation.
Using the total number of exposures, $n_{tot} = n_{ex} T / t_{p}$, 
and the linear pixel size, $p_{x} \sim \lambda f / a S$, we have:

\[ \sigma_{m} \ \sim \ \sigma_{smp} \ \sqrt{\frac{1}{n_{tot}}}
  \ \frac{p_{x}}{d}  \]

The product of $\ \sigma_{smp}$ and $p_{x}$ is nothing else but the
single measure precision expressed in linear units ($\mu$m), which
we call $\sigma_{sml}$, thus

\[ \sigma_{m} \ \sim \ \frac{\sigma_{sml}}{d} \ \sqrt{\frac{1}{n_{tot}}} \]

\subsection{Discussion}

From the above assumptions and the result for the best achievable 
astrometric mission precision, $\sigma_{m}$ for the stare-mode 
concept we see the following:

\begin{enumerate}
\item The value for $\sigma_{m}$ does NOT depend on the aperture of 
the telescope, nor the focal length, nor the sampling.
If we want to shift the ``sweet spot" to a required magnitude,
this of course needs to be accomplished by a certain combination
of aperture, exposure time, throughput (bandwidth, quantum efficiency...)
and mission lifetime. 
However, the numeric value for the best astrometry remains
unaffected by shifting it to a desired magnitude, thus is 
independent of aperture, focal length, bandwidth and QE.

\item The $\sigma_{m}$ does not depend on wavelength.
Thus we are free to choose the spectral regime we want the
mission to operate in.  The choice of a specific wavelength
will require a match between the focal length and pixel size with 
the desired sampling.  This will not affect the $\sigma_{m}$.
However, manufacturability and alignment tolerances are
better met for a red than blue spectral bandpass.
For a given, linear tolerance (fixed cost), the wave front error 
as a fraction of the wavelength is smaller for red than for
blue light which buys an advantage in image quality.

\item The $\sigma_{m}$ does not depend on the pixel size directly,
however it does depend on the product of pixel fraction error and
pixel size, i.e.~depends directly on the linear measure precision
in the focal plane.
For very small pixels the limiting factor will be the physical
structures in the pixels, for very large pixels the limiting 
factor will be the pixel fraction for the image centroiding.
It is important to minimize $\sigma_{sml}$.

\item The driving factors toward smaller $\sigma_{m}$ are
a large number of single measurements and a large focal plane.
The large number of measures (with a constant mission life time)
implies many, short exposures with minimal overhead.  This is
somewhat contrary to the basic mission concept, and particularly 
with a requirement to have many visits, as for example for the
science goals to detect many exo-planets.  
However, a very large number of observations then heavily relies 
on the $\sqrt{n}$ law, which fails at some point due to 
systematic errors.

\item The biggest and easiest impact on $\sigma_{m}$ can be
achieved by increasing the linear size of the focal plane. 
Smaller astrometric errors can be obtained by spending more
money in the focal plane than for the aperture of the
telescope optics.  For sampling near critical 
($\approx$ 2 pixel/FWHM) and for visible to near-infrared
wavelengths the f-ratio of the optical system will be slow
(order f/30).   This is an advantage for an optics design.
The focal length and the angular size of the field of view
will only affect the number of visits, not $\sigma_{m}$ directly.

\end{enumerate}

\section{Examples}

Table 3 gives numerical values for 2 example stare-mode missions, 
a large aperture mission, like the current OBSS baseline, and a small,
feasibility-study-type mission like the Milli-arcsecond 
Pathfinder Survey (MAPS) satellite currently under study at USNO
\citep{maps,aasmp}.  MAPS will also be a technology demonstration mission,
using a single, large-format CMOS or CMOS-hybrid detector.
 Even the small MAPS mission would
be capable of improving over Hipparcos by a factor of about 3
in positional precision and a factor of 100 in number of stars.
The MAPS-like example given in Table 3 gives about 250 $\mu$as
mission precision for a single, well exposed star, while the
stated MAPS goal is to achieve at least 1 mas accuracy.
For the MAPS mission the required extragalactic targets would
be near the limiting magnitude of the general all-sky survey
with unfavorable S/N ratio.  However the flexibility of the
stare-mode concept would permit observation of the required number
of relatively faint targets with more observations and longer
integration times.

The magnitudes at the bottom of Table 3 for the CMOS case
assume that high precision stellar image centroids can be
obtained for stars up to 3.0 magnitudes brighter than the
saturation limit.  No charge bleeding will be present and
such bright stellar images can be fitted using the unsaturated
wings of the profile.
This is a conservative estimate, good astrometric results
might be obtained for even brighter stars.  
The optical quality of the observed point spread function
on the 1\% level of the peak intensity and below, 
straylight and other factors
will limit the astrometric quality of such centroids at
some point if a stellar image is vastly overexposed even
if the detector does not bleed at all.

Expected results from the large OBSS-type mission are very much
comparable to Gaia, with the additional benefit of being able
to reach fainter stars in a general all-sky survey and
going significantly deeper for targeted fields.
For a comprehensive discussion of OBSS capabilities and science
goals see the NASA roadmap report by USNO \citep{obss},
where also other issues of concern are mentioned together
with suggested solutions.
The shutter issue, the required quality of the guiding, 
CPU power requirements, downlink rate and several other 
issues of possible concern are not intractable problems.

Systematic errors will likely limit the performance of
any astrometric mission.
Comparing the 2 FOV, large-angle measurement approach
with the single FOV, differential mode, both have to 
cope with imperfections of the optics and image centroiding
issues.  Imperfect knowledge of field distortions,
shifts of centroid positions as a function of magnitude
and color of the stars will affect both types of missions
similarly, while performing accurate, differential
measures in the focal plane.
A scanning mission might have some advantage because the 
signal for each observation is averaged over many pixels.
The 2 FOV approach has the disadvantage of absolute,
large-angle measures with the basic angle stability
problem, a possible source of significant additional 
systematic errors.

\section{Conclusions}

For a Hipparcos-type mission the scanning operation concept was
a good choice.  With no access to a sufficient number of extragalactic
targets, large-angle measures (2 fields of view separated by order 
of $90^{\circ}$) are essential to obtain absolute parallaxes.
Even today, if an astrometric mission would be limited to 12th
magnitude a Hipparcos-type concept would be the way to go.
For a mission capable of accessing extragalactic sources and
being able to move between fields fast as compared to the
integration time, the stare-mode concept becomes a viable 
alternative to the traditional 2-FOV scanning operation.
Both conditions are met for the stare-mode missions under
consideration now (OBSS and MAPS).
For these missions, a 2-FOV scanning concept would still have 
the advantage of many individual observations, which is important 
for some science goals like detecting extra-solar planets.
The achievable astrometric mission precisions are comparable
between the scanning and stare-mode concept, if the large
square-root-n factor for a scanning mission can be believed
for the mission accuracy estimate.
However, the stare-mode concept is a lower-risk approach with a 
high single measure precision, a more conservative square-root-n
factor and simpler engineering requirements, thus lower costs.
It has the advantage that no technology developments are needed.

The major advantages of a stare-mode astrometric mission are
the high degree of flexibility, the higher astrometric precision
in targeted areas, the ability to go significantly deeper and
the reduced complexity of the hardware with easier to achieve
engineering requirements.
The stare-mode concept also potentially allows the use of 
radiation-hard CMOS detectors which have an inherent large 
dynamic range (no blooming) thus also would allow low-risk 
access to relatively bright stars, particularly in combination 
with short exposure times.  Exposure times in general are 
unrestricted in stare-mode operations, contrary to the scanning 
mode of operation.

At this time, the stare-mode concept is not nearly as well 
developed as the scanning satellite concept and detailed 
simulations will soon be performed to better understand 
the exact requirements, capabilities and limitations.
A particularly appealing aspect of a stare-mode mission
like OBSS is that it has the ability to be complementary
or a replacement mission depending on the observing schedule
but using the same hardware.
If Gaia performs as predicted, a stare-mode mission can take 
the Gaia reference frame and concentrate on deep, targeted fields.
If Gaia does not perform as currently envisioned, the stare-mode 
mission independently can fulfill most of Gaia's science goals 
in a mainly general all-sky survey.
Furthermore, because the design is fundamentally different than
Gaia, it offers redundancy in case of problems in the Gaia
implementation.
A large enough stare-mode mission could also substitute for many
of the SIM targeted mission science.
In order to get a real proof-of-concept, a small, 
feasibility-study-type mission like MAPS is strongly suggested, 
which could have a fast turnaround time and give valuable insight
for the planning of a full-scale mission.



\acknowledgments
The authors thank Ralph Gaume, Hugh Harris, Ken Johnston and Sean Urban for 
valuable discussions and comments on the draft version of this paper.
The referee is thanked for many comments which lead to an improved paper.

\clearpage

\begin{table}
\begin{center}
\caption{Basic input parameters which define the mission for the
  astrometric mission precision analysis.}
\begin{tabular}{lcl}
\tableline
\tableline
Item Description                  & Symbol      & Unit   \\
\tableline
number of exposures per pointing  & $n_{ex}$    &        \\
mean exposure time single expos.  & $t_{ex}$    &  s     \\
overhead time per pointing        & $t_{ov}$    &  s     \\
total mission time                & $T$         &  year  \\
single measure precision          & $\sigma_{smp}$ & pixel \\
\tableline
aperture                    & $a$         &  m     \\
focal length                & $f$         &  m     \\
sampling                    & $S$         & pixel/FWHM \\
diameter of focal plane     & $d$         &  m     \\
central wavelength          & $\lambda$   &  nm    \\
\tableline
\end{tabular}
\end{center}
\end{table}

\clearpage

\begin{table}
\begin{center}
\caption{Derived parameters characterizing a stare-mode mission
  leading to the astrometric mission precision for well-exposed stars.}
\begin{tabular}{lrcll}
\tableline
\tableline
Item Description        &       &  & Dependence  & Unit   \\
\tableline
resolution              & $r$   &=& 1.22 $\lambda$ \ 0.206265 / $a$ & mas \\ 
FWHM of image profiles  & FWHM  &$\approx$& 1.0 \ r  & mas \\
image scale             & $s $  &=& 206.265 / $f$     & ``/mm = mas/$\mu$m\\
diameter of field of view  & FOV   &=& $d$ \ $s$ \ 1000 / 3600 & degree \\
pixel scale             & $p_{s}$  &=& FWHM / $S$        & mas / pixel  \\
linear pixel size       & $p_{x}$  &=& $p_{s}$ / $s $       & $\mu$m       \\
integr.~time per pointing & $t_{i}$ &=& $n_{ex} t_{ex}$ & s            \\
time spend at 1 pointing& $t_{p}$ &=& $t_{i} + t_{ov}$& s            \\
tot.numb. single observ.& $n_{tot}$ &=& $n_{ex}$ \ 86400 \ 365.25 \ T / \ $t_{p}$ & \\
sky area per pointing   & $A$   &$\approx$& 0.7 \ FOV$^{2}$ & square degree\\
time for 1 sky coverage & $t_{s}$ &=& (41,000 / $A$) $t_{p}$ / 86400 & day \\
observing frequency     & $\nu $ &=& 365.25 / $t_{s}$ & visits/year   \\
numb.single obs.per star& $n   $ &=& T \ $\nu $ \ $n_{ex}$ &        \\
angular single measure error & $\sigma_{1}$ &=& $\sigma_{smp}$ \ $p_{s}$ &mas\\
mission precision       & $\sigma_{m}$ &=& $\sigma_{1}$ / $\sqrt{n}$ & mas\\
\tableline
\end{tabular}
\end{center}
\end{table}

\clearpage

\begin{table}
\begin{center}
\caption{Example missions: a 1.5-meter aperture OBSS and a small test satellite (MAPS)}
{\footnotesize
\begin{tabular}{llrr}
\tableline\tableline
Item  & Unit & OBSS  &  MAPS \\
\tableline
number of exp.~per pointing &     &       2 &       2 \\
tot.overh.time per pointing & [s] &    20.0 &    10.0 \\
size single CCD, linear  &[mm] &    50.0 &    80.0 \\
total mission time       &[yr] &     5.0 &     2.5 \\
single meas.precision  &[$\mu$m] & 0.050 &   0.100 \\
fraction of gaps between CCDs& &   0.150 &   0.000 \\
\tableline
aperture                &  [m] &    1.50 &    0.15 \\
pixel size          & [$\mu$m] &    10.0 &    10.0 \\
sampling        & [pixel/FWHM] &     2.5 &     2.0 \\
number of CCDs           &     &     360 &       1 \\
long  exposure time      & [s] &    20.0 &    15.0 \\
short exposure time      & [s] &     2.0 &     5.0 \\
central wavelength      & [nm] &   600.0 &   650.0 \\
overall QE(system)& [fraction] &    0.60 &    0.60 \\
width of bandpass       & [nm] &    300. &    200. \\
read noise              & [e$^{-}$] & 7.0 &   15.0 \\
\tableline
tot. numb. pixel        &[Gpx] &    9.00 &    0.06 \\
resolution 1.22 lam/ap  &[mas] &   100.7 &  1090.5 \\
FWHM profile = 1.0  res &[mas] &   100.7 &  1090.5 \\
pixel scale          &[mas/px] &    40.3 &   545.2 \\
scale             & [mas/$\mu$m] &  4.03 &   54.52 \\
focal length            &  [m] &   51.23 &    3.78 \\
f-number  apert/foc.length &   &    34.2 &    25.2 \\
foc. plane area px.cover.&[m$^{2}$] & 0.900 &0.006 \\
focal plane diameter     & [m] &    1.15 &    0.09 \\
diameter FOV        & [degree] &    1.28 &    1.37 \\
\tableline
tot.integr.time per pointing& [s] &    22.0 &    20.0 \\
sky area per pointing  & [sq.deg] &    1.43 &    1.87 \\
time for all sky once  & [day] &    13.9 &     7.6 \\
observ. frequency &[visits/yr] &    26.3 &    48.0 \\
tot.numb. single obs.~per star& &  262.7 &   239.8 \\
best single meas.error & [mas] &   0.201 &   5.452 \\
best single meas.err&[1/pixel] &    200. &    100. \\
sqrt(n) best mission prec.&[$\mu$as] &    12.4 &   352.1 \\
... at limit. mag       &[$\mu$as] &    351. &   4980. \\
\tableline
full well capacity      &[ke-] &    123. &    123. \\
limiting mag. short exp.&[mag] &    18.4 &    14.0 \\
limiting mag. long  exp.&[mag] &    20.9 &    15.2 \\
\tableline
saturation mag.short ex.&[mag] &    11.6 &     7.6 \\
saturation mag. long ex.&[mag] &    14.4 &     8.7 \\
faint limit best astrom.&[mag] &    15.4 &    11.0 \\
range of mags best astr.&[mag] &     1.0 &     2.3 \\
\tableline
CMOS saturat. short exp.&[mag] &     8.6 &     4.6 \\
CMOS saturat. long  exp.&[mag] &    11.4 &     5.7 \\
CMOS range best astrom. &[mag] &     4.0 &     5.3 \\
\tableline
\end{tabular}
}
\end{center}
\end{table}


\end{document}